\documentstyle[12pt,prl,aps]{revtex}

\draft
\begin{document}
\title{
Cluster Algorithm for hard spheres and related systems}
\author{Christophe Dress$^*$ and Werner Krauth$^{**,\dagger}$}
\address{
$^{*}$ CNRS-Laboratoire de Physique Th\'{e}orique de l'ENS\\
24, rue Lhomond; F-75231 Paris Cedex 05; France\\
e-mail: dress@physique.ens.fr\\
$^{**}$ CNRS-Laboratoire de Physique Statistique de l'ENS\\
24, rue Lhomond; F-75231 Paris Cedex 05; France\\
e-mail: krauth@physique.ens.fr\\
}
\date{September 1995}
\maketitle
\begin{abstract}
In this paper, we present a cluster algorithm for the 
simulation of  hard spheres and related systems. 
In this algorithm, a copy of the  configuration is 
rotated with respect to a randomly chosen pivot point.
The two systems are then superposed, and clusters of overlapping spheres
in the joint system are isolated. Each of these clusters can 
be ``flipped'' independently, a process which generates 
non-local moves in the original configuration. 
A generalization of this algorithm
(which works perfectly well at small density) can be successfully
made to work at densities around the solid-liquid transition point in 
the two-dimensional hard-sphere system. 
\end{abstract}
$^{\dagger}$ Corresponding author
\newpage
Since the 1987 paper by Swendsen and Wang~\cite{swendsen} 
(and the subsequent paper
of Wolff~\cite{wolff}), the simulation of 
Ising or $XY$-type systems close to the critical point has been 
much simplified: just as the physical system, 
conventional Monte Carlo algorithms (for a review cf. \cite{binder}) suffer from 
critical slowing down, but the new algorithms overcome this
problem and allow the calculation of {\em thermodynamic} quantities
with great ease.

One of the long-standing problems in classical statistical physics is
the hard-sphere liquid~\cite{alder57}. 
In two dimensions, the transition between
the liquid and the solid order in the  hard-sphere liquid has been
the subject of unabating interest~\cite{alder62}.
There  are several competing theoretical scenarios for the transition,
and Monte Carlo work has been going on for more than $30$ years~\cite{halperin}
(for a review {\em cf.} \cite{strandburg}).
However, the conventional local-move Monte
Carlo simulations are greatly affected by the slowing down
of the simulation around the transition. 
At present, the maximum size of the simulation box, which 
can be unequivocally thermalized, contains only of the order of $1000$ 
particles~\cite{lee}.  Much larger simulations have been 
undertaken~\cite{zollweg,marx} and 
sophisticated data analysis has been performed~\cite{marx}. 
However, due to the 
fact that the probability distribution has not yet converged
to its equilibrium value, these simulations are biased
in a way which is very difficult to assess.

In this paper, we present a cluster algorithm, which is applicable
to the hard sphere system in any dimension, and which is easily
generalized to incorporate an additional potential.
The main idea of the algorithm is to rotate a copy of the ``current''
configuration, and to superpose this rotated copy with the original
simulation box. Clusters are then isolated in the {\em joint} system.
Each of the clusters is then flipped independently, {\em i. e.} the
spheres belonging to a cluster are moved from the rotated
copy into the original configuration and vice versa.
For concreteness, consider the fig. 1, which illustrates the 
algorithm:
the original configuration $c_1$ (fig. 1a) is made up of 
$N$ spheres of radius $r$
in a box of size ($L_x, L_y$) (in the figure $N=9$).
In addition to the original configuration, we
consider also a configuration $c_2$ (displayed in fig. 1b), which is 
obtained from $c_1$ by a $\pi$-rotation:
We generate $c_2$ by picking 
an arbitrary ``pivot'' point $p=(p_x,p_y)$ with
$0<p_x<L_x, 0<p_y<L_y$ (In the example, $p_x=0.52 \;L_x  , 
p_y= 0.53 \;  L_y$).
We then rotate $c_1$ around $p$ of an angle $\pi$ to obtain $c_2$~\cite{footperiod}.
The choice of the pivot is the essential Monte Carlo element of the algorithm.

The two configurations $c_1$ and $c_2$ are then superposed
as shown in fig. 1c,  where they form clusters of overlapping 
spheres~\cite{footcluster}.
Two types of clusters are possible:
``even'' clusters,  made up of an equal number of spheres in $c_1$ and 
$c_2$, and ``odd'' clusters, in which the numbers differ. In fig. 1c, 
{\em e. g.}, the cluster ``II'' is even, while cluster ``I'' is odd.

Generally, clusters appear in pairs (such as I and 
IV), 
with the possible exception of a single ``even'' cluster which is
symmetric around the pivot  (III).

It is now easily seen that we may ``flip'' the clusters, {\em i. e.} 
interchange between $c_1$ and $c_2$ the spheres belonging to a cluster.
We are interested in performing a canonical simulation in which
the individual numbers of
spheres  both of  $c_1$ and $c_2$ have to remain unchanged.  We therefore
choose to perform such flips 
for individual even clusters, or for pairs of odd clusters,
such as I and IV. 
The result of one such cluster flip (of 
clusters I and IV) is shown in fig. 1d. 
Finally, we restrict our attention back to the updated configuration,
$c_1'$
in the original simulation box.
By inspection of fig. 1e, which shows $c_1'$, we see that picking $p$ and
flipping clusters I and IV has achieved
a {\em nonlocal} Monte Carlo move: 
spheres $8$ and $9$ were moved 
from the lower left corner to the upper right one,
and the sphere $6$ from the upper right corner to the lower left one.
 
It should be evident that - given an arbitrary pivot point - 
the flip satisfies 
the detailed balance condition, and constitutes a viable Monte Carlo
move. To see this one simply needs to consider the ``reverse'' move (from 
fig. 1e back to  fig. 1a), which has exactly the same probability
to occur as the original one. 

Applying the same argument as above to an even cluster
(such as V in fig. 1c), we notice that small {\em even} clusters
generate only {\em local} moves. Due to the limited benefits of 
worrying about (small) even clusters, we usually exclude them 
from our considerations.
Many generalizations are possible: It is evident that the 
spheres can have different radii {\em etc}; a potential can 
be taken into account in usual way, by calculating the 
Boltzmann weights of the proposed flip and the reverse one;
furthermore, the angle of the rotation around the pivot can be chosen 
at will. This only introduces some effects far from the pivot, which can 
be eliminated.

The simple algorithm which has just been described works perfectly well.
At small density, 
the  combined system of $c_1$ and $c_2$ breaks up into a large number
of small clusters, which can be flipped independently.
At higher density ({\em i. e.} above the percolation threshold
of this combined system), there is a single percolating cluster
(which it is useless to flip), and an {\em algebraically decaying}
distribution of small clusters~\cite{stauffer}.
Just as in the Swendsen-Wang algorithm,
there is a ``magical'' point, the percolation threshold (which, for
the Ising model, corresponds to the Curie temperature~\cite{kasteleyn}). 
At this point,
the behavior of the algorithm is optimal. In our
case, the ``magical point'' is the percolation threshold of the 
system of {\em superposed} configurations, which unfortunately lacks
physical interest.
In two dimensions, we find this percolation
threshold to be  situated at a density 
of  $\rho \sim 0.62$, definitely lower than the densities in 
which we are interested 
(as usual~\cite{lee}, the density is defined as the ratio of the
number of spheres and the volume of the simulation box, $\rho = N/V$,
normalized to $2/\sqrt{3}$ for the most compact state; in these 
units, the transition takes place around $\rho \approx 
0.9$~\cite{alder62,marx}).
Around the percolation threshold, the algorithm 
decorrelates the whole system by flipping a few large cluster, 
as in the Wolff algorithm~\cite{wolff}.

Close to the liquid-solid
transition in the two-dimensional system,  {\em i. e.} much above the
percolation point, it is particularly difficult to 
find a sufficient number of small {\em odd} clusters,
which generate the non-local moves.
It is difficult to find odd clusters, but one rather often 
encounters configurations which {\em almost} constitute odd clusters, as
the ones presented in fig. 2, which are kept from flipping by a few
weak ''links``. We now present a stochastically correct trick
which has allowed up to break up a large number of these weak links.

For fixed but arbitrary $\epsilon$, we define an $\epsilon-$cluster
as a set of spheres which may have an arbitrary number of $\epsilon-$links,
{\em i. e.} links between a disk of the
set and a disk of the boundary (in a different box), 
larger than $2 \times r -\epsilon$ (so that the overlap between the 
spheres is smaller than $\epsilon$). 
In addition to having $\epsilon-$links, the 
$\epsilon-$cluster itself 
is held together by links which are not $\epsilon-$links, {\em i. e.}
which are shorter than 
$2 \times r - \epsilon$. 

After 
isolating an $\epsilon-$cluster, we ``freeze'' the boundary,
and perform a certain fixed 
number of $n_{loc}$ {\em local} Monte Carlo moves exclusively
of the spheres in
the $\epsilon-$cluster (without destroying it). Each time
the number of $\epsilon-$links falls to zero, we have obtained
a true cluster, which we flip. It is easily shown that this exotic
``dynamic of $\epsilon-$clusters'' satisfies detailed balance, 
since we make sure that at each step
both the initial and the final configuration are $\epsilon-$clusters.

We have programmed the complete algorithm sketched in this paper. For 
small systems (up to $14$ spheres), we performed  extremely long
runs both  of a standard MC algorithm and of the present one. We find
identical probability distributions {\em e. g.} for the orientational
order parameter~\cite{leefoot} to a precision of $0.1 \%$. There is thus very 
little room for  doubt about the correctness of 
the present algorithm, and for programming errors in our actual 
implementation.

It is easily understood that, at high density, the main workload of
the algorithm consists in the determination of the percolating cluster.
Since we never actually ``flip'' this cluster, most of the effort is
thus spent in finding out what one does not  want to do - 
a frustrating  way of using CPU time . Only after discarding the 
percolating cluster,  we have a chance of finding small 
odd clusters. 
A moment's thought suffices to understand
that there is a  faster way to find the small odd clusters. Consider
the cluster IV the fig. 1c): it is evident that the nonlocal move
can be performed only under the condition that, locally, it is 
possible to replace the sphere 6 by two other spheres. Whether
there is at all enough space in a given neighborhood  to replace $n$ 
spheres by $n+1$ can (for small $n=0, 1,2$) be decided by a {\em local}
analysis which, in addition,  is {\em independent} of the pivot 
(this remark is pertinent to clusters and $\epsilon-$clusters).
An improved algorithm of the kind presented in this paper thus 
first isolates the loci at which $n$ spheres may be replaced 
by $n+1$ (for small $n=0, 1,2$). Once this analysis is done, 
one chooses randomly pivots, and then does the - now trivial - cluster search, 
in regions which have survived the screening stage. 
Since the local analysis just described can be reused a large
number of times (and updated, once we have flipped a cluster, or
$\epsilon-$cluster), the cluster search is much simplified. 
The final algorithm is thus quite efficient in generating 
non-local moves.

Let us finally stress that the method presented in this paper is very
general, and may be applied to a large number of systems. The
specific application to the hard-sphere liquid stands out as
prototypical, and we hope that the algorithm will be helpful in
elucidating the order of the transition, in the measure of correlation
functions {\em etc}. Work along these lines is in progress.

\acknowledgments
We acknowledge a very helpful discussion with D. Marx.

\noindent
\newpage
{\bf Figure Captions}

\begin{enumerate}
\item
\label{move}
The algorithm presented in this paper performs nonlocal moves
( a) $\rightarrow$ e) by considering a randomly rotated copy ( b) )
of the actual configuration ( a)). a) and b) are superposed ( c) )
and clusters are isolated ( c)) and flipped ( d)) in the superposed
configuration.
\item
\label{epsilon}
We present a trick which allows to flip not only clusters (as in fig.
\ref{move}), but also $\epsilon-$clusters.
\end{enumerate}

\begin{references}
\bibitem{footperiod} Periodic 
boundary conditions and the possible proximity of the cluster to the pivot have
to be treated carefully.
\bibitem{footcluster} All the  clusters in fig. 1c 
can be determined in a total of $O(N)$ operations.
\end{references}
\end{document}